\documentclass[seceq]{ptptex}

\usepackage{graphicx}

\def \ketv #1>{\mbox{$|{#1}\rangle$}} 
\def \mate<#1|#2|#3>{\mbox{$\langle {#1}|\,{#2}\,|{#3}\rangle$}}
\def \etal{{\it et al.\ }}
\def \ie{{\it ie.\ }}

\def\YTD(#1,#2){%
\unitlength=6pt%
\begin{picture}(#1,2)(0,0)%
\multiput(0,0)(1,0){#2}{\framebox(1,1)}%
\multiput(0,1)(1,0){#1}{\framebox(1,1)}%
\end{picture}}

\def\YTT(#1,#2,#3){%
\unitlength=6pt%
\begin{picture}(#1,3)(0,1)%
\multiput(0,0)(1,0){#3}{\framebox(1,1)}%
\multiput(0,1)(1,0){#2}{\framebox(1,1)}%
\multiput(0,2)(1,0){#1}{\framebox(1,1)}%
\end{picture}}

\title{
A pentaquark model of the negative parity $\Lambda(1405)$
}

\author{
Takashi \textsc{Inoue}\footnote{ e-mail address:
takash-i@sophia.ac.jp}  
}

\inst{
Sophia University,
Tokyo 102-8554, Japan
}

\abst{%
We test a pentaquark model of the negative parity hyperon $\Lambda(1405)$.
We use the model where valence four quarks and one antiquark move in a potential relativisticaly 
and interact with themselves and each other through gluon and NG boson.
We calculate mass shift of the system and find that 
there are two $\Lambda(1405)$ with almost degenerate mass:
one flavor singlet and one flavor octet. 
We calclate also $\sigma$-term and scalar form factor of the hyperon
and show their usefulness to deduce the number of valence quark in the hypeon.
}

\begin{document}

\maketitle

\section{Introduction}

The negative parity hyperon $\Lambda(1405)S_{01}$
is interested for many years and several interpretations are proposed.
The most simplest one is so called P-wave excited 3-quark system.
Isgur and Karl studied such system with harmonic-oscillator wave function
and one-gluon-exchange perturbation. 
While, Furuichi \etal studied it as a bound state problem. 
It turns out that it is difficult to reproduce mass of $\Lambda(1405)$ in this approach.
The most popular scenario today is bound $\bar K N$ system 
or dynamically generated resonance in meson-baryon scattering. 
In this decade, mason-baryon coupled channel scattering are studied extensively.
There, several baryons including $\Lambda(1405)$, are generated dynamically
as a resonance or a quasi bound state which decay to open channels.
Most of studies based on the chiral Lagrangian and the Bethe-Salpeter equation.
A lattice study shows that SU(3) flavor singlet 3-quark system
with $I(J^P) = 0(1/2^-)$, cannot be so light as $\Lambda(1405)$. 

Besides these scenario, pentaquark is a possible interpretation of the hyperon.
Here, pentaquark stands for a state with five valence constituents.
In other words, it's quark number decomposition starts form $\ketv qqqq \bar q>$ states.
By today, we do not have established pentaquark.
Indeed $\Theta^+$ is a good candidate though it's existence is still questionable.
Therefore, if $\Lambda(1405)$ is well described as a pentaquark, 
we encounter exotic pentaquark baryon for the first time.

The $\pi N$ $\sigma$-term is defined by
$
 \sigma_{\pi N}\!=\!\langle N| \! \int \!\! d^3 \vec x ~ 
                 m_u \, \bar u u(\vec x) + m_d \, \bar d d(\vec x) \, |N\rangle
$
is a measure of violation of the chiral symmetry in the real world,
and gives quark mass contribution to nucleon mass roughly.
Important point is that $\sigma$-terms are very sensitive to number of quark
involved in baryon, including sea-quark. 
In fact, sea-quark is so significant that it contribute more than 65\%
of the empirical value of the $\pi N$ $\sigma$-term. 
The same nucleon matrix element but with finite momentum transfer
is called scalar form factor $\sigma_{\pi N}(Q^2)$.

We test the pentaquark picture of $\Lambda(1405)$.
We study mass, $\sigma$-term and scalar form factor of $\Lambda(1405)$. 
Any $\sigma$-terms can be defined as the same as $\pi N$ $\sigma$-term.

\section{Pentaquark model}%

To make negative parity baryon with four quarks and one anti-quark,
it is simplest to put all quarks and antiquark into the ground S-wave orbit, 
namely S-shell pentaquark. 
The whole system should be color singlet for QCD quark confinement.
On top of that, 4-quark subsystem must be totally antisymmetric from the Fermi statistics. 
There are several flavor-spin configuration satisfying these conditions.
We consider three configurations where 4-quark subsystem is
\begin{equation}
 (\, {\bf [2,1,1]_f} ~,~ \bf{ [3,1]_s } \,) \quad \mbox{and} \quad%
 (\, {\bf [2,1,1]_f} ~,~ \bf{ [2,2]_s } \,) \quad \mbox{and} \quad%
 (\, {\bf [2,2]_f  } ~,~ \bf{ [3,1]_s } \,) %
\end{equation}
in flavor and spin respectively.
With $\bf[211]_f$ subsystem, pentaquark is flavor singlet with an antiquark.
While, $\bf [22]_f$ one leads flavor octet and anti-decuplet pentaquarks.
A $\Lambda$ hyperon is at center of octet.
With $\bf [31]_s$ subsystem, pentaquark can be both spin parity $J^P = 1/2^-$ and $3/2^-$. 
We chose former one for $\Lambda(1405)$.
In all, we have three pentaquark $\Lambda(1405)$, two flavor single and one octet.
In conventional 3-quark approach, $\Lambda(1405)$ is made as flavor singlet,
while in the chiral unitary model, one pole corresponding to $\Lambda(1405)$
is almost flavor octet\cite{Jido:2002yz}.
At this stage, there is no specific quark correlation and no extra quarks in pentaquark.  


We consider valence quark localized by a potential $S(r) - \gamma^0 V(r)$, 
whose field is expanded in terms of bound states $u_{\alpha}(x)$ and $v_{\beta}(x)$.
We define unperturbed pentaquark states as a product of the ground states,
\begin{equation}
 \ketv \mbox{\small Penta} >^0 = u_0(x_1) u_0(x_2) u_0(x_3) u_0(x_4) \bar v_0(x_5)
\end{equation}
and introduce a residual interaction by which valence quarks/antiquark emit and absorb
perturbative gluon and pseudoscalar meson octet(NG boson)
\begin{equation}
 {\cal L}_I = - \bar{\psi}(x) \biggl\{ 
     g_s \gamma^\mu A_\mu^a(x) \frac{\lambda^a}{2} +
     S(r) i \gamma^5 \frac{\hat \Phi (x)}{F} 
                \biggr\}\psi(x)
\end{equation}
where the quark meson coupling term is made so that it recover the chiral symmetry
broken by the scalar potential $S(r)$. 
We set up perturbation where baryon matrix element of a local operator is calculated
using the unperturbed state and the interaction.
There, the interaction generates quark correlation and
meson cloud supplementing baryon \ie extra quarks in baryon.
This formalism is developed by T\"ubingen group for nucleon originally, 
and called the perturbative chiral quark model(PCQM)\cite{Lyubovitskij:2000sf}.


First, we study mass spectrum of three $\Lambda(1405)$ in the present pentaquark picture.
Baryon mass is shifted by two origins in the present model:
valence quark mass and the residual interaction.
The former mass shift is given by
\begin{eqnarray}
& & \delta M_{\Lambda(1405)} \simeq
      \hat m   \, \gamma   \, \left(\#u + \#d + \#\bar u + \#\bar d\right)   
         + m_s \, \gamma_s \, \left(\#s + \#\bar s \right)   
\\ 
& & \qquad
 \begin{array}{ccccccc}
 \hline
                & \#u & \#d  & \#s & \#\bar u & \#{\bar d} & \#\bar s \\
 \Lambda(1405,1_f) & 4/3   & 4/3  & 4/3 & 1/3      & 1/3        & 1/3 \\
 \Lambda(1405,8_f) & 3/2   & 3/2  & 1   & 1/2      & 1/2        & 0 \\
 \hline
 \end{array}
\end{eqnarray}
with quark masses $\hat m$, $m_s$ and relativistic reduction factor $\gamma$, $\gamma_s$.
As you see, the flavor octet $\Lambda$ pentaquark has advantage compared to the singlet one.
While, mass shift by the residual interaction is evaluated perturbatively as
\begin{equation}
\hspace{-17pt}
\delta M_{\Lambda(1405)} \simeq
 {}^0\!\langle \Lambda | {\sum_{n=1}^{\infty}} \frac{i^n}{n!} 
   \! \int \! i \delta(t_1) d^4 x_1 \ldots d^4 x_n  
   T[ {{\cal L}_I (x_1) \ldots {\cal L}_I(x_n)} ]
                  | \Lambda \rangle^0_c
\end{equation}
but we truncate it at second order of the interaction.

Next, we study the following $\sigma$-term of $\Lambda(1405)$, 
namely $\pi \Lambda(1405)$ $\sigma$-term.
\begin{equation}
 \sigma_{\pi \Lambda(1405)} 
  = \hat m  \langle \Lambda(1405)|
 \! \int \! \! d^3 \vec x 
 \left\{ S_u^{\mbox{\tiny PCQM}}(\vec x) + S_d^{\mbox{\tiny PCQM}}(\vec x)
 \right\} |\Lambda(1405) \rangle
\end{equation}
Here,  $S_i^{\mbox{\tiny PCQM}}(\vec x)$ is the scalar density operator relevant to $\sigma$-term,
which is derived from the model Hamiltonian instead of the QCD one, and given by 
\begin{equation}
  S_u^{\mbox{\tiny PCQM}}(x) 
  = \frac{ \partial {\cal H}^{\mbox{\tiny PCQM}}(x) }{\partial m_u}
  = \bar u u(x) 
      + B \left\{ \pi^+ \pi^- + \frac{\pi^0 \pi^0}{2} 
                      + K^+ K^- + \frac{\eta\eta}{6} \right\}
\end{equation}
for example\cite{Lyubovitskij:2000sf}. 
The first term corresponds to valence quark, 
while the second term does meson cloud \ie sea-quark.
Where, $B$ is the low energy constant in the standard $\chi$SB scenario.
The matrix element is evaluated perturbatively. 

\section{Results and discussions}

\begin{table}[t]
\caption{Mass shifts and $\sigma$-term of pentaquark $\Lambda(1405)$.}
\label{tbl:massandsigma}
\begin{center}
\begin{tabular}{l|ccc|ccccc}
\hline \hline
                           & \multicolumn{3}{|c}{$\delta M_{\Lambda(1405)}$ [MeV]} 
                           & \multicolumn{5}{|c}{$\sigma_{\pi \Lambda(1405)}$ [MeV]} \\
                           & $\hat m, m_s$ & gluon  & meson 
                           &  val &  $\pi$ &  $K$   &  $\eta$  &  Total \\
\hline
 $\Lambda(1405,{8_f})$            & $+159$ & $-572$ & $-434$ & $20.5$  & $31.5$ & $2.3$ & $0.24$ & $54.5$ \\
 $\Lambda(1405,{1_f},S_{4q}={1})$ & $+248$ & $-629$ & $-489$ & $17.1$  & $30.2$ & $3.5$ & $0.45$ & $51.3$ \\
 $\Lambda(1405,{1_f},S_{4q}={0})$ & $+248$ & $-528$ & $-410$ \\ 
\hline \hline
\end{tabular}
\end{center}
\end{table}

For numerical evaluation of diagrams, 
we use standard value of quark masses $\hat m = 7$ MeV, $m_s = 25 \hat m$,
and low energy constants $B = 1.4$ GeV and $F = 88$ MeV.
We use a potential localizing valence quark of linear $S(r) = c r$ with $c=0.11$ GeV$^2$
and a constant $V(r)$, for trial.
Where, proton charge radius is 0.76 [fm] at tree level,
and the reduction factors are $\gamma=0.73$ and $\gamma_s=0.79$.  
For internal quark line, we use propagator in the potential
but omit excited states for simplicity, and z-diagram to avoid double counting with meson.
For gluon propagator, we use dressed one with the running coupling of the form by Maris and Robers
but low momentum part is fitted to $N$-$\Delta$ mass difference.

Table \ref{tbl:massandsigma} center column shows obtained mass shifts.
As noted before, concerning valence quark mass,
the flavor octet one has lower mass of 89 MeV compared to the singlet one.
While, regarding to interaction,
the flavor singlet one with spin one 4-quark subsystem is energetically favored
by 112 MeV compared to the octet one.
The another singlet one with spin zero 4-quark subsystem
is not favored energetically at all compared to others.
Therefore, we have two $\Lambda(1405)$ with almost degenerate masses in this model:
one singlet and one octet.
This point agrees with a consequence of the chiral unitary model\cite{Jido:2002yz}.

In the present model, the $\pi N$ $\sigma$-term is obtained as
$
 \sigma_{\pi N} = 15.4  + 34.5 + 0.95 + 0.03  = 50.9
$ 
MeV with valence quark and $\pi$, $K$, $\eta$-induced sea-quark contributions respectively, 
We see that it well agrees with empirical value
and pion induced sea-quark is most important.
Table \ref{tbl:massandsigma} right column shows obtained $\pi\Lambda$ $\sigma$-term.
We see that it is almost same as the $\pi N$ one.
Since number of u, d valence quark in octet one is 4 instead of 3, 
the valence contribution increase by factor 4/3. 
While, number of pion loop is a little less than nucleon case, 
and the decrease of pion contribution cancel the increase of valence one.
Number of $K$ and $\eta$ diagram is much more than nucleon case
due to valence $s$ quark in $\Lambda$ hyperon. 
But, their magnitudes are tiny and do not increase the total $\sigma$-term much.
This large $\sigma$-term more than 50 MeV clearly
shows that the present pentaquark is more than $\ketv qqqq\bar q>$.

\begin{figure}[t]
\begin{center}
 \includegraphics[width=0.45\textwidth]{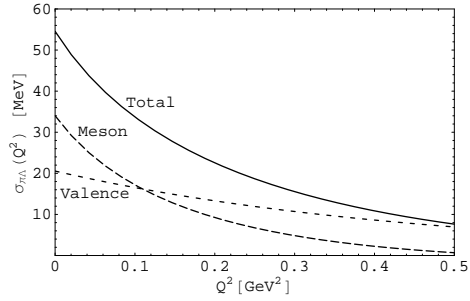}
 \caption{Obtained scalar form factor of $\Lambda(1405)$.}
 \label{fig:scffpilam}
\end{center}
\end{figure}

Fig. \ref{fig:scffpilam} shows obtained $\pi \Lambda(1405)$ scalar form factor,
where valence and meson partial contributions are also shown\cite{Inoue:2003bk}. 
We see that meson cloud or sea-quark contribution disappear rapidly at finite $Q^2$,
and the form factor at $Q^2 > 0.5$ [GeV$^2$] is dominated by valence quark contribution.
This means that we can deduce information about number of valence quark
by seeing scalar form factor at finite recoil.
For example, the present pentaquark picture predict
that the ratio $\sigma_{\pi \Lambda(1405,8_f)}(Q^2)/\sigma_{\pi N}(Q^2)$ at medium recoil is 4/3
due to the difference in number of valence quark. 
Although such data is not available today,
this could be used to distinguish pictures of $\Lambda(1405)$ in future.
For this aim, we need to invent way to access $\sigma$-term of the hyperon experimentally.
Although we've studied only ${\pi\Lambda}$ $\sigma$-term in this paper,
${K \Lambda}$ or ${\eta \Lambda}$ $\sigma$-term might be easier to access 
and/or more useful to study nature of the hyperon.

\section*{Acknowledgements}
The authors thanks the Yukawa Institute for Theoretical Physics at Kyoto University, 
where this work was completed during the YKIS2006 on "New Frontiers on QCD". 
This work is supported by the Grant for Scientific Research No.18042006 
from the Ministry of Education, Culture, Science and Technology, Japan.

\appendix


\end{document}